\documentclass{article}
\usepackage{euscript,amsmath,amssymb,amsfonts,graphicx,bm,amsthm,mathtools}
  \usepackage{caption}
\usepackage{subcaption}
\usepackage{pdflscape}

\usepackage{nameref} 

\usepackage{array}

\usepackage{cite} 

  \usepackage{paralist}
  \usepackage{graphics} 
\usepackage{soul}

\usepackage{latexsym,amssymb,enumerate,amsmath,amsthm} 
  \usepackage{graphicx} 
  \usepackage{tikz}
 \usepackage[hidelinks]{hyperref}

\usepackage{latexsym,amssymb,enumerate,amsmath} 
\usepackage{pgfplots}
\pgfplotsset{compat=1.18}
\usepackage{soul,xcolor}

\usepackage{lineno}
\newcommand*\patchAmsMathEnvironmentForLineno[1]{%
  \expandafter\let\csname old#1\expandafter\endcsname\csname #1\endcsname
  \expandafter\let\csname oldend#1\expandafter\endcsname\csname end#1\endcsname
  \renewenvironment{#1}%
     {\linenomath\csname old#1\endcsname}%
     {\csname oldend#1\endcsname\endlinenomath}}%
\newcommand*\patchBothAmsMathEnvironmentsForLineno[1]{%
  \patchAmsMathEnvironmentForLineno{#1}%
  \patchAmsMathEnvironmentForLineno{#1*}}%
\AtBeginDocument{%
\patchBothAmsMathEnvironmentsForLineno{equation}%
\patchBothAmsMathEnvironmentsForLineno{align}%
\patchBothAmsMathEnvironmentsForLineno{flalign}%
\patchBothAmsMathEnvironmentsForLineno{alignat}%
\patchBothAmsMathEnvironmentsForLineno{gather}%
\patchBothAmsMathEnvironmentsForLineno{multline}%
}

\theoremstyle{plain}
\theoremstyle{remark}

\theoremstyle{definition}

\newcommand{\dd}{\textup{d}}
\def\eps{\varepsilon}

\def\P{\mathbb{P}}
\def\R{\mathbb{R}}

\newcommand{\Markov}[2]{\underset{#1}{\overset{#2}{\rightleftharpoons}}}
\newcommand{\Dl}{D_{\mathrm{w}}}
\newcommand{\Dr}{D_{\mathrm{e}}}
\newcommand{\Db}{D_{\mathrm{b}}}
\newcommand{\Dc}{D_{\mathrm{c}}}
\newcommand{\alphal}{\alpha_{\mathrm{w}}}
\newcommand{\alphar}{\alpha_{\mathrm{e}}}
\newcommand{\rhol}{\rho_{\mathrm{w}}}
\newcommand{\rhor}{\rho_{\mathrm{e}}}
\newcommand{\gb}{\gamma_{\mathrm{w}}}
\newcommand{\gc}{\gamma_{\mathrm{c}}}
\newcommand{\gbr}{\gamma_{\mathrm{e}}}
\newcommand{\X}{\mathbf{X}}
\newcommand{\x}{\mathbf{x}}

\begin{document}


\title{Channel transport: gating, geometry, and heterogeneous diffusion}


\author
{Sean D. Lawley\thanks{Department of Mathematics, University of Utah, Salt Lake City, UT 84112 USA (\texttt{lawley@math.utah.edu}).}
}
\date{\today}
\maketitle

\begin{abstract}
Channel-mediated transport is ubiquitous in biology. A series of works by different theoreticians have sought to determine how the diffusive flux through a channel depends on (a) stochastic gating, (b) channel geometry, and (c) heterogeneous diffusion. In this paper, we derive an explicit estimate for the diffusive flux through a channel that accounts for these three factors. We show that our estimate is exact in certain parameter regimes. We further use stochastic simulations to confirm that our estimate remains accurate across a very broad range of parameters. Our estimate differs from some results in the physics literature.
\end{abstract}

\tableofcontents

\section{\label{sec:introduction}Introduction}

Diffusive transport through channels or pores is ubiquitous in biology. For example, cells use membrane channels to transport ions between the interior and exterior of the cell \cite{phillips2012physical}. Channels also facilitate transport between subcellular compartments, such as nuclear pores, which mediate exchange between the nucleus and cytoplasm of eukaryotic cells \cite{kozai2025karyopherins}. An example with channels on a larger spatial scale  is insect respiration, which relies on diffusive transport of oxygen and carbon dioxide through a network of tracheal tubes \cite{wigglesworth31}. Understanding and quantifying the flux in these and other biological systems is challenging due to three important features:
\begin{enumerate}
    \item[\hspace{1cm}(a)] stochastic gating,
    \item[\hspace{1cm}(b)] channel geometry, and
    \item[\hspace{1cm}(c)] different diffusivity inside versus outside the channel.
\end{enumerate}

\begin{figure}
	\centering
	\includegraphics[width=1\textwidth]{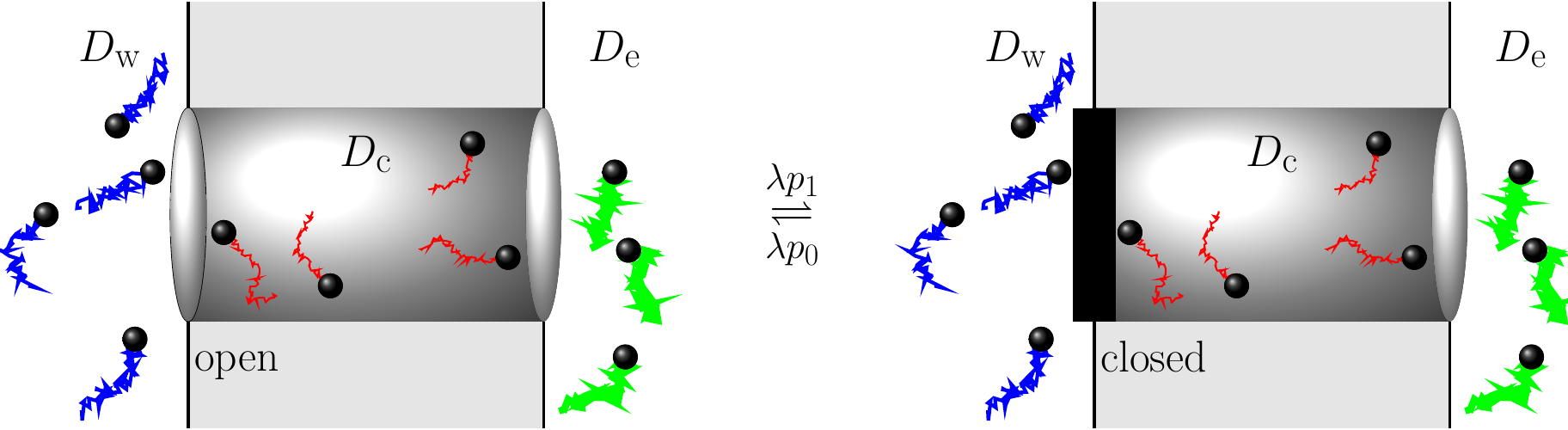}
	\caption{Diffusing particles can pass from the left (or ``west'') bulk region through a channel to the right (or ``east'') bulk region. Particles have diffusivities $\Dl$, $\Dc$, and $\Dr$ in the west bulk, channel, and east bulk. The west entrance to the channel is stochastically gated, which means that it randomly opens and closes. Particles can freely enter and exit the west side of the channel when the gate is open (left diagram), whereas particles can neither enter nor exit through the west side of the channel when the gate is closed (right diagram).}
	\label{fig:schem}
\end{figure}

Elaborating on (a), the channel entrance in many biological systems stochastically opens and closes. In the open state, diffusing ``particles'' (ions, molecules, proteins, etc.)\ can freely enter and exit the channel through the gate, whereas the closed state restricts particles from entering or exiting through the gate. See Figure~\ref{fig:schem} for an illustration. Stochastic gating is often modeled by a two-state Markov jump process,
\begin{align}\label{eq:markov0}
    \text{open}\Markov{\lambda p_0}{\lambda p_1}\text{closed},
\end{align}
where $p_0=1-p_1$ denotes the fraction of time that the gate is open and $\lambda$ parameterizes the overall switching rate between open and closed states.

How does gating affect the flux through the channel? It is often assumed that the average flux through a gated channel, $J$, is simply the product of the flux through a channel that is always open, $J^{\text{open}}$, and the fraction of time that the gate is open,
\begin{align}\label{eq:naive}
    J
    =p_0 J^{\text{open}}.
\end{align}
Indeed, the simple relation in \eqref{eq:naive} is a fundamental assumption of the Hodgkin-Huxley model and other conductance-based neuron models \cite{hodgkin1952, piccolino2002, hausser2000}. However, the relation in \eqref{eq:naive} was challenged in Refs.~\cite{lawleythesis, lawley15sima}, which analyzed a simple model of gated transport in a one-dimensional channel of length $L$ and diffusivity $\Dc$. Refs.~\cite{lawleythesis, lawley15sima}  argued that \eqref{eq:naive} only holds if the switching rate is much slower than the timescale of diffusion in the channel, i.e.\ if $\gamma_{\text{c}}:=\sqrt{\lambda L^2/\Dc}\ll1$. In fact, Refs.~\cite{lawleythesis, lawley15sima} argued that the gated flux is identical to the ``always open'' flux if the switching rate is fast,
\begin{align}\label{eq:fast0}
    \lim_{\gamma_{\text{c}}\to\infty}J
    =J^{\text{open}}.
\end{align}
The prediction in \eqref{eq:fast0} is counterintuitive because it means that even if the gate is open only a small fraction of time ($p_0\ll1$), the gated flux can be nearly as high as the flux through a channel that is always open. 
The counterintuitive prediction in \eqref{eq:fast0} has been contradicted by more recent models and analyses in the physics literature \cite{berezhkovskii2016diffusive, berezhkovskii2017effect, berezhkovskii2018effect, berezhkovskii2018stochastic}. 

Elaborating on (b), though the flux is understood to be greater through wider and shorter channels, precisely quantifying how channel geometry controls the flux is more difficult. Mathematically, determining this geometric control entails analyzing the diffusion equation in a non-trivial three-dimensional geometry. Indeed, the one-dimensional approximation used in some prior works \cite{lawleythesis, lawley15sima} may break down if the channel is not sufficiently narrow.  

Elaborating on (c), the particle diffusivity may change upon entering the channel. For example, ATP is thought to diffuse around 10 times slower in voltage-dependent anion channels compared to free solution \cite{rostovtseva1997vdac}. Mathematically, such space-dependent diffusivity or ``heterogeneous diffusion'' introduces multiplicative noise and the associated It\^o versus Stratonovich controversy \cite{mannella2012ito} and requires accounting for the so-called ``spurious'' force associated with any non-It\^o interpretation \cite{serov2020statistical, vaccario2015}. 

In this paper, we formulate and analyze a model of diffusive transport through a channel that connects two ``bulk'' reservoirs. See Figure~\ref{fig:schem} for an illustration. Our main result is an explicit approximation for the average flux through the channel that accounts for (a) stochastic gating, (b) channel geometry, and (c) a diffusivity that is different in the channel versus the bulks. If the channel is a cylinder of radius $a$ and length $L$ and is stochastically gated according to \eqref{eq:markov0}, then our estimate of the flux of particles which enter the channel from the ``west'' (left) bulk and ultimately wander far away into the ``east'' (right) bulk is
\begin{align}\label{eq:J0}
    J_*
    =4a\Dl c_\infty\mathcal{P}_*,
\end{align}
where $\Dl$ is the diffusivity in the west bulk, $c_\infty$ is the fixed particle concentration far away in the west bulk, and 
\begin{align*}
    &\mathcal{P}_*=
    \bigg[\frac{4\rhol}{\pi} \Big(1+\frac{p_1}{p_0}\frac{\tanh (\gc)}{\gc }+\frac{\pi}{4\rhor}+\frac{(p_1/p_0) \text{sech}^2(\gc)}{ \rhor (\frac{4}{\pi}+\gbr)+\gc  \tanh (\gc)}\Big)+\frac{\frac{4}{\pi}+\gb p_0}{\frac{4}{\pi} p_0+\gb p_0}\bigg]^{-1},
\end{align*}
which is the probability that a particle which reaches the west entrance to the channel from far away in the west bulk will eventually traverse the channel and wander far away into east bulk. The probability $\mathcal{P}_*$ is a function of the fraction of time that the gate is open, $p_0=1-p_1$, the following three dimensionless parameters,
\begin{align*}
    \gb=\sqrt{\frac{\lambda a^2}{\Dl}},\quad
    \gc=\sqrt{\frac{\lambda L^2}{\Dc}},\quad
    \gbr=\sqrt{\frac{\lambda a^2}{\Dr}},
\end{align*}
which compare the switching rate $\lambda$ to the diffusion timescales in the west bulk, channel, and east bulk (where $\Dc$ and $\Dr$ are the respective diffusivities in the channel and east bulk), and the following two dimensionless parameters,
\begin{align*}
    \rhol=\frac{L}{a}\bigg(\frac{\Dl}{\Dc}\bigg)^{\alphal},\quad
    \rhor=\frac{L}{a}\bigg(\frac{\Dr}{\Dc}\bigg)^{\alphar},
\end{align*}
which compare the channel aspect ratio $L/a$ to the change in diffusivity at the west and east channel entrances, where $\alphal\in[0,1]$ and $\alphar\in[0,1]$ specify the interpretation of the multiplicative noise (where the values $\alpha=0$, $\alpha=1/2$, and $\alpha=1$ correspond respectively to the It\^o \cite{ito1944stochastic}, Stratonovich \cite{stratonovich1966new}, and kinetic interpretations \cite{hanggi1982stochastic}). We derive \eqref{eq:J0} by analyzing a system of reaction-diffusion equations in three spatial dimensions. We also generalize \eqref{eq:J0} to channels with more general cross-sectional shapes.

We show that our flux estimate $J_*$ in \eqref{eq:J0} is exact if $\min\{\rhol,\rhor\}\gg1$ in the sense that
\begin{align}\label{eq:exactasymptotic0}
    J_*
    /J_{\text{exact}}\to1\quad\text{as }\min\{\rhol,\rhor\}\to\infty,
\end{align}
where $J_{\text{exact}}$ is the exact flux (which is not known explicitly). We further use stochastic simulations to show that $J_*$ is accurate in a very broad range of parameters,
\begin{align*}
    J_*
    \approx J_{\text{exact}}. 
\end{align*}
Indeed, we have been unable to find a parameter regime in which $J_*$ is inaccurate. We also show that $J_*$ reduces to some prior flux estimates \cite{lawleythesis, lawley15sima, lawley2026diffusive} in certain special cases and that $J_*$ satisfies the counterintuitive relation for fast switching in \eqref{eq:fast0} if we take $\min\{\gb,\gc,\gbr\}\to\infty$.  We show that $J_*$ differs from some prior flux estimates derived in the physics literature \cite{berezhkovskii2017effect, berezhkovskii2018effect, berezhkovskii2018stochastic}.

The rest of the paper is organized as follows. We start in section~\ref{sec:alwaysopen} by deriving the following flux estimate for the case that the channel is always open,
\begin{align*}
    J_*^{\text{open}}
    =4a\Dl c_\infty\mathcal{P}_*^{\text{open}},\quad\text{where }\mathcal{P}_*^{\text{open}}=\frac{1}{1+\rhol/\rhor+(4/\pi)\rhol}.
\end{align*}
We then consider a gated channel in section~\ref{sec:gating2} and derive $J_*$ in \eqref{eq:J0}. Section~\ref{sec:asymptotics} explores the behavior of $J_*$ in various asymptotic regimes, including showing that $J_*$ is exact in the sense of \eqref{eq:exactasymptotic0}. Section~\ref{sec:simulations} compares $J_*$ to stochastic simulations. We conclude in section~\ref{sec:discussion} by discussing related work, including discrepancies between $J_*$ and prior estimates. 

\section{\label{sec:alwaysopen}Always open channel}

\subsection{Setup}

Consider a particle diffusing in an unbounded, three-dimensional domain given by the union $\Omega_{\text{w}}\cup\Omega_{\text{c}}\cup\Omega_{\text{e}}\subset\R^3$, where the left bulk, channel, and right bulk are the following three respective sets,
\begin{align*}
    \Omega_{\text{w}}
    &=\{(x,y,z)\in\R^3:x<-L/2\},\\
    \Omega_{\text{c}}
    &=\{(x,y,z)\in\R^3:x\in(-L/2,L/2),\,(y,z)\in a\Gamma\},\\
    \Omega_{\text{e}}
    &=\{(x,y,z)\in\R^3:x>L/2\}.
\end{align*}
The subscripts ``w,'' ``c,'' and ``e'' are for ``west,'' ``channel'' (or ``central''), and ``east.'' The channel $\Omega_{\text{c}}$ has length $L>0$, diameter $\mathcal{O}(a)$, and cross-sectional shape $a\Gamma\subset\R^2$, where $\Gamma\subset\R^2$ is a dimensionless shape with on order one diameter $\text{diam}(\Gamma)=\mathcal{O}(1)$ (recall that the diameter of a set is the maximum possible distance between any two points in that set). We are most interested in the case that the channel is a cylinder of radius $a$ (in which case $\Gamma$ is the unit disk), but we proceed in terms of a general cross-sectional shape.

Suppose that the diffusivity is the following piecewise constant function of the $x$ coordinate,
\begin{align}\label{eq:Dx}
    D(x,y,z)=D(x)
    =\begin{cases}
        \Dl & \text{if }x<-L/2,\\
        \Dc & \text{if }x\in(-L/2,L/2),\\
        \Dr & \text{if }x>L/2.
    \end{cases}
\end{align}
To specify the interpretation of the multiplicative noise (inherent to any space-dependent diffusivity \cite{van1981ito, mannella2012ito, sokolov2010ito, vaccario2015, tung2025escape, tung2026stochastic}), suppose that the probability density $p(x,y,z,t)$ that the particle is at position $(x,y,z)$ at time $t$ satisfies the following continuity conditions at the west channel entrance,
\begin{align*}
    \Dl^{1-\alphal}p(-L/2^-,y,z,t)
    &=\Dc^{1-\alphal}p(-L/2^+,y,z,t),\quad (y,z)\in a\Gamma,\\
    \Dl\partial_xp(-L/2^-,y,z,t)
    &=\Dc\partial_xp(-L/2^+,y,z,t),\quad (y,z)\in a\Gamma,   
\end{align*}
and analogous conditions at the east channel entrance,
\begin{align*}
    \Dc^{1-\alphar}p(L/2^-,y,z,t)
    &=\Dr^{1-\alphar}p(L/2^+,y,z,t),\quad (y,z)\in a\Gamma,\\
    \Dc\partial_xp(L/2^-,y,z,t)
    &=\Dr\partial_xp(L/2^+,y,z,t),\quad (y,z)\in a\Gamma,
\end{align*}
where $g(x_0^\pm)=\lim_{x\to x_0^\pm}g(x)$ denotes one-sided limits. The parameters $\alphal\in[0,1]$ and $\alphar\in[0,1]$ specify the interpretation of the multiplicative noise\cite{bressloff2017temporal}, with the most common \cite{mannella2012ito}  interpretations being $\alpha=0$ (It\^{o} \cite{ito1944stochastic}) and $\alpha=1/2$ (Stratonovich \cite{stratonovich1966new}). Another convention is $\alpha=1$, which is sometimes called kinetic, isothermal, or H{\"a}nggi-Klimontovich \cite{hanggi1982stochastic}.

The flux $J$ of particles which enter the channel from the west bulk and ultimately wander far into the east bulk is the product of (i) the flux of particles from the west bulk which reach the west entrance to the channel at $x=-L/2$, and (ii) the probability $\mathcal{P}$ that a particle which arrives at the west channel entrance from the west bulk will eventually wander far into the east bulk and never return to the channel. That is, the flux $J$ is given by
\begin{align}\label{eq:JP}
        J
    =(2\pi C_0(\Gamma)a\Dl c_\infty)\mathcal{P},
\end{align}
where $C_0(\Gamma)$ is the electrostatic capacitance of $\Gamma$ (capacitance is defined so that the capacitance of a unit disk is $2/\pi$). The geometric prefactor $2\pi  C_0(\Gamma)a$ can be derived via a simple  divergence theorem argument \cite{lawley2026escape}. Note that $2\pi C_0(\Gamma)a=4a$ for a cylindrical channel\cite{berg1977}.

\subsection{Splitting probability}

Suppose that the stochastic gate is always open. 
Let $P^{\text{open}}(\x)=P(\x)$ be the probability that a particle that starts at $\x=(x,y,z)$ will eventually wander far away into the east bulk and never return to the channel,
\begin{align*}
    P(\x)
    :=\P\Big(\limsup_{t\to\infty}X(t)\ge L/2\,\Big|\,\X(0)=\x\Big),
\end{align*}
where $(X(t),Y(t),Z(t))$ denotes the stochastic position of the particle at time $t$. 
The function $P$ is harmonic in the two bulks and in the channel,
\begin{align}\label{eq:Pharmonic}
    \Delta P
    &=0,\quad \x\in\Omega_{\text{w}}\cup\Omega_{\text{c}}\cup\Omega_{\text{e}},
\end{align}
satisfies the far-field conditions,
\begin{align*}
    \lim_{\|\x\|\to\infty,\,x<-L/2}P(\x)
    &=0,\quad
    \lim_{\|\x\|\to\infty,\,x>L/2}P(\x)
    =1,
\end{align*}
continuity conditions at $x=-L/2$, 
\begin{align}\label{eq:contentrance}
        P(-L/2^-,y,z)
    &=P(-L/2^+,y,z)\quad (y,z)\in a\Gamma,\\
    \Dl^{\alphal}\partial_xP(-L/2^-,y,z)
    &=\Dc^{\alphal}\partial_xP(-L/2^+,y,z)\quad (y,z)\in a\Gamma,
\end{align}
and analogous continuity conditions at $x=L/2$,
\begin{align}
\begin{split}\label{eq:contentranceL}
    P(L/2^-,y,z)
    &=P(L/2^+,y,z)\quad (y,z)\in a\Gamma,\\
    \Dc^{\alphar}\partial_xP(L/2^-,y,z)
    &=\Dr^{\alphar}\partial_xP(L/2^+,y,z)\quad (y,z)\in a\Gamma,
\end{split}    
\end{align}
and satisfies reflecting boundary conditions on the rest of the boundary of $\Omega_{\text{w}}\cup\Omega_{\text{c}}\cup\Omega_{\text{b}}$.

For a particle which reaches the west channel entrance from far away in the west bulk, the probability of wandering far into the east bulk is
\begin{align}\label{eq:mathcalP2}
    \mathcal{P}(-L/2)
    &:=\int_{a\Gamma} P(-L/2,y,z)\,\mu(y,z)\,\dd y\,\dd z,
\end{align}
where $\mu(y,z)$ is the probability density of where the particle hits the channel entrance if it started far from the channel and is conditioned to hit the channel. That is, $\mu(y,z)$ is the so-called harmonic measure density\cite{grebenkov2015analytical}. If the channel is a cylinder of radius $a$, then $\mu$ has a density given by the following explicit function\cite{sneddon1966mixed} of the radius $r=\sqrt{y^2+z^2}\in[0,a)$,
\begin{align}\label{eq:hittingdensity}
    \mu(r)&=\frac{r}{2\pi a^2\sqrt{1-(r/a)^2}},\quad r^2=y^2+z^2<a^2.
\end{align}

Let $\overline{P}$ denote the uniform average of $P$ in the channel cross-section,
\begin{align}
    \overline{P}(x)
    &:=\frac{1}{|a\Gamma|}\int_{a\Gamma}P(x,y,z)\,\dd y\,\dd z,
\end{align}
where $|a\Gamma|=a^2|\Gamma|$ denotes the area of the channel cross-section.
Integrating \eqref{eq:Pharmonic}, using the two-dimensional divergence theorem and the reflecting boundary conditions, and solving the resulting ordinary differential equations yields that $\overline{P}(x)$ is the following linear function,
\begin{align*}
    \overline{P}(x)
    &=\overline{P}(-L/2)+(x/L)\big(\overline{P}(L/2)-\overline{P}(-L/2)\big).
\end{align*}
Taking the derivative yields
\begin{align}\label{eq:Pderiv}
    \overline{P}'(-L/2^+)
    =\big(\overline{P}(L/2)-\overline{P}(-L/2)\big)/L>0.
\end{align}

\subsection{Two exact equations}

We now derive two exact equations involving the four probabilities $\mathcal{P}(\pm L/2)$ and $\overline{P}(\pm L/2)$. Integrating \eqref{eq:Pharmonic} over a large hemisphere $H_R$ of radius $R\gg a$ in the left bulk,
\begin{align}\label{eq:hemisphere}
    H_R
    =\{(x,y,z)\in\R^3:x<-L/2,\,(x+L/2)^2+y^2+z^2<R^2\},
\end{align}
and using the divergence theorem yields
\begin{align}\label{eq:inthemi}
    0
    =\int_{H_R}\Delta P\,\dd \x
    =\int_{\partial H_R^+}\partial_{\mathbf{n}}P\,\dd S
    +\int_{a\Gamma}\partial_x P(-L/2^-,y,z)\,\dd y\,\dd z,
\end{align}
where $\partial_{\mathbf{n}}$ denotes the normal derivative and $\partial H_R^+$ denotes the curved part of the boundary of the hemisphere $H_R$,
\begin{align}\label{eq:boundaryofhemisphere}
    \partial H_R^+
    =\{(x,y,z)\in\R^3:x<-L/2,\,(x+L/2)^2+y^2+z^2=R^2\}.
\end{align}
The strong Markov property implies that $P$ has the following monopole decay at far-field\cite{lawley2026escape},
\begin{align*}
    P(\x)
    \sim \frac{C_0(a\Gamma)\mathcal{P}(-L/2)}{\|\x\|}\quad\text{as }\|\x\|\to\infty,\,x<-L/2,
\end{align*}
and therefore
\begin{align*}
    \int_{\partial H_R^+}\partial_{\mathbf{n}}P\,\dd S
    \to
    -2\pi aC_0(\Gamma)\mathcal{P}(-L/2)<0
    \quad\text{as }R\to\infty.
\end{align*}
Furthermore, the second continuity condition in \eqref{eq:contentrance} and the derivative of $\overline{P}$ in \eqref{eq:Pderiv} imply
\begin{align*}
    \int_{a\Gamma}\partial_x P(-L/2^-,y,z)\,\dd y\,\dd z
    &=\Big(\frac{\Dc}{\Dl}\Big)^{\alphal}\int_{a\Gamma}\partial_x P(-L/2^+,y,z)\,\dd y\,\dd z\\
    &=\Big(\frac{\Dc}{\Dl}\Big)^{\alphal}|\Gamma|a\frac{a}{L}(\overline{P}(L/2)-\overline{P}(-L/2)).
\end{align*}
Therefore,
\begin{align}\label{eq:exact}
    \mathcal{P}(-L/2)
    =\frac{1}{G(\Gamma)}\frac{(\overline{P}(L/2)-\overline{P}(-L/2))}{\rhol},
\end{align}
where $G(\Gamma)$ is the following dimensionless geometric factor,
\begin{align}\label{eq:G}
    G(\Gamma)
    =\frac{2\pi C_0(\Gamma)}{|\Gamma|}>0,
\end{align}
and $\rhol$ is the following dimensionless comparison of the tube aspect ratio and the change of diffusivity at the west channel entrance, 
\begin{align}\label{eq:rhodeftube}
    \rhol
    :=(L/a)(\Dl/\Dc)^{\alphal}>0.
\end{align}
Note that the geometric factor in \eqref{eq:G} is $G(\Gamma)=4/\pi$ if the tube is a cylinder (i.e.\ if $\Gamma$ is the unit disk).

By symmetry, we obtain the analog of \eqref{eq:exact} at the east channel entrance,
\begin{align}\label{eq:exactright}
    1-\mathcal{P}(L/2)
    =\frac{1}{G(\Gamma)}\frac{(\overline{P}(L/2)-\overline{P}(-L/2))}{\rhor},
\end{align}
where $\mathcal{P}(L/2)$ is defined analogously to \eqref{eq:mathcalP2},
\begin{align*}
    \mathcal{P}(L/2)
    &:=\int_{a\Gamma} P(L,y,z)\,\mu(y,z)\,\dd y\,\dd z,
\end{align*}
and $\rhor$ is defined analogously to \eqref{eq:rhodeftube},
\begin{align*}
    \rhor
    :=(L/a)(\Dr/\Dc)^{\alphar}>0.
\end{align*}
We emphasize that \eqref{eq:exact} and \eqref{eq:exactright} are exact relations.

\subsection{Approximation}

To obtain an explicit approximation from the exact relations \eqref{eq:exact} and \eqref{eq:exactright}, we ignore variation in the channel cross-section and make the approximations
\begin{align}\label{eq:ignorecrosssectionalvariation}
    \mathcal{P}(-L/2)
    \approx \overline{P}(-L/2),\quad
    \mathcal{P}(L/2)
    \approx \overline{P}(L/2),
\end{align}
so that \eqref{eq:exact} and \eqref{eq:exactright} becomes a linear system in the two variables $\mathcal{P}(\pm L/2)$, which we solve to obtain the approximation
\begin{align}\label{eq:Popen}
    \mathcal{P}(-L/2)
    &\approx\mathcal{P}_{*}^{\text{open}}
    :=\frac{1}{1+\rhol/\rhor+G(\Gamma)\rhol},
\end{align}
where the superscript ``open'' emphasizes that \eqref{eq:Popen} is for the case that the channel is always open. The formula \eqref{eq:Popen} generalizes a result of Ref.~\cite{lawley2026escape} to the case that the diffusivities in the two bulks may differ.

We show below that the approximation \eqref{eq:Popen} is exact if $\rhol\gg1$ and $\rhor\gg1$. We note that if $\rhol=\rhor=\rho$ (which happens if $\Dl=\Dr$ and $\alphal=\alphar$ or $\alphal=\alphar=0$), then it was shown in Ref.~\citenum{lawley2026escape} using stochastic simulations that \eqref{eq:Popen} is accurate for all $\rho>0$ in the case of a cylindrical channel. We further note that $\lim_{\rhor\to\infty}\mathcal{P}_*^{\text{open}}=1/(1+G(\Gamma)\rhol)$, which was an approximation derived in Ref.~\citenum{richardson2026yet} for the probability of merely reaching the east end of the tube (rather than reaching the east end of the tube and then wandering far away into the east bulk).

\subsection{Approximation is exact as $\min\{\rhol,\rhor\}\to\infty$}

We claim that the approximation $\mathcal{P}_*^{\text{open}}$ in \eqref{eq:Popen} is exact if $\rhol\gg1$ and $\rhor\gg1$. That is, we claim
\begin{align}\label{eq:showexact}
    \mathcal{P}(-L/2)
    \sim\frac{1}{G(\Gamma)\rhol}
    \sim\mathcal{P}_*^{\text{open}}
    \quad\text{as }\min\{\rhol,\rhor\}\to\infty,
\end{align}
where $f\sim g$ means $f/g\to1$. We need only show the first asymptotic equivalence in \eqref{eq:showexact}, since the second asymptotic equivalence in \eqref{eq:showexact} is immediate from \eqref{eq:Popen} (we assume throughout this paper that $G(\Gamma)=\mathcal{O}(1)$; note that $G(\Gamma)=4/\pi\approx1.27$ if the channel is a cylinder). The first asymptotic equivalence in \eqref{eq:showexact} follows from the exact relation in \eqref{eq:exact} if we can show that
\begin{align}
    \lim_{\rhol\to\infty}\overline{P}(-L/2)
    &=0,\label{eq:firstvanish}\\
    \lim_{\rhor\to\infty}\overline{P}(L/2)
    &=1.\label{eq:secondvanish}
\end{align}
By the strong Markov property,
\begin{align}\label{eq:bound}
    \overline{P}(-L/2)\le \overline{u}(-L/2),
\end{align}
where $\overline{u}(x)=\frac{1}{a^2|\Gamma|}\int_{a\Gamma}u(x,y,z)\,\dd y\,\dd z$ and $u(x,y,z)$ is the probability that the diffusing particle will reach the east end of the channel given that it started at $(x,y,z)$. The limit $\lim_{\rhol\to\infty}\overline{u}(-L/2)=0$ was shown in Ref.~\cite{richardson2026yet}, and thus \eqref{eq:bound} implies \eqref{eq:firstvanish}. The result \eqref{eq:secondvanish} holds by symmetry.

\section{\label{sec:gating2}Gated channel}

\subsection{Setup}

Consider the model of section~\ref{sec:alwaysopen}, but suppose that there is a stochastic gate at the west channel entrance that opens and closes according to the Markov jump process,
\begin{align}\label{eq:markov}
    \text{open}\quad 0\Markov{\lambda_1}{\lambda_0}1\quad\text{closed},
\end{align}
where the opening and closing rates are
\begin{align}
\begin{split}\label{eq:lambda01}
    \lambda_1
    &=p_0\lambda\quad\text{(opening rate)},\\
    \lambda_0
    &=p_1\lambda\quad\text{(closing rate)},
\end{split}    
\end{align}
and $p_0$ and $p_1=1-p_0$ are the steady-state probabilities of the gate being open and closed,
\begin{align}
\begin{split}\label{eq:stationary}
    p_0
    &=\frac{\lambda_1}{\lambda_0+\lambda_1}\quad\text{(probability open)},\\
    p_1
    &=\frac{\lambda_0}{\lambda_0+\lambda_1}\quad\text{(probability closed)},
\end{split}    
\end{align}
and $\lambda=\lambda_0+\lambda_1$ parameterizes the ``switching rate'' between the open and closed states.

\subsection{Splitting probability}

Let $P_0(\x)$ (respectively, $P_1(\x)$) be the probability that a particle that starts at position $\x=(x,y,z)$ will eventually wander far into the east bulk and never return to the channel, given that the gate is initially open (respectively, closed). These probabilities satisfy the following Kolmogorov backward equations,
\begin{align}
\begin{split}\label{eq:backward}
    0
    &=D(x)\Delta P_0-\lambda_0(P_0-P_1),\\
    0
    &=D(x)\Delta P_1-\lambda_1(P_1-P_0),
\end{split}    
\end{align}
where $D(x)$ is in \eqref{eq:Dx}, with far-field conditions,
\begin{align*}
    \lim_{\|\x\|\to\infty,\,x<-L/2}P_0(\x)=\lim_{\|\x\|\to\infty,\,x<-L/2}P_1(\x)=0,\\
    \lim_{\|\x\|\to\infty,\,x>L/2}P_0(\x)=\lim_{\|\x\|\to\infty,\,x>L/2}P_1(\x)=1.
\end{align*}
At the west entrance to the channel, $P_0$ satisfies the continuity conditions in \eqref{eq:contentrance} and $P_1$ satisfies no flux conditions,
\begin{align}\label{eq:noflux}
    \partial_x P_1(-L/2^-,y,z)
    =\partial_x P_1(L/2^+,y,z)
    =0,\quad (y,z)\in a\Gamma.
\end{align}
At the east entrance to the channel, $P_0$ and $P_1$ satisfy the continuity conditions in \eqref{eq:contentranceL}. 

Diagonalize \eqref{eq:backward} by introducing
\begin{align}
    P
    &=p_0P_0+p_1P_1,\label{eq:Pdefn}\\
    Q
    &=P_0-P_1\label{eq:Qdefn},
\end{align}
so that $P$ and $Q$ satisfy the following decoupled partial differential equations,
\begin{align}
    \Delta P
    &=0,\label{eq:Pbackward}\\
    D(x)\Delta Q
    &=\lambda Q.\label{eq:Qbackward}
\end{align}
Furthermore, $P$ and $Q$ satisfy the far-field conditions,
\begin{align*}
    \lim_{\|\x\|\to\infty,\,x<-L/2}P(\x)
    &=\lim_{\|\x\|\to\infty,\,x<-L/2}Q(\x)=0,\\
    \lim_{\|\x\|\to\infty,\,x>L/2}P(\x)
    &=1,\quad \lim_{\|\x\|\to\infty,\,x>L/2}Q(\x)=0,
\end{align*}
and $P$ and $Q$ satisfy the continuity conditions at the east end of the channel in \eqref{eq:contentranceL}. 
The functions $P$ and $Q$ are decoupled, except for their conditions at the west end of the channel, 
which can be derived from the fact that $P_0$ satisfies \eqref{eq:contentrance}, $P_1$ satisfies \eqref{eq:noflux}, and the following representations (which are equivalent to \eqref{eq:Pdefn}-\eqref{eq:Qdefn}),
\begin{align*}
    P_0
    &=P+p_1Q,\\
    P_1
    &=P-p_0Q.
\end{align*}

If a particle arrives at the west entrance to the channel from far away in the west bulk, then the gate will be in its stationary distribution ($p_0$ and $p_1$ in \eqref{eq:stationary}), and thus
\begin{align}\label{eq:mathcalPfull}
    \mathcal{P}(-L/2^-)
    =p_0 \mathcal{P}_0(-L/2)
    +p_1 \mathcal{P}_1(-L/2^-),
\end{align}
where
\begin{align}
    \mathcal{P}_0(-L/2)
    &=\int_{a\Gamma} P_0(-L/2,y,z)\,\mu(y,z)\,\dd y\,\dd z,\label{eq:mathcalPfull0}\\
    \mathcal{P}_1(-L/2^-)
    &=\int_{a\Gamma} P_1(-L/2^-,y,z)\,\mu(y,z)\,\dd y\,\dd z,\label{eq:mathcalPfull1}
\end{align}
and $\mu$ is the harmonic measure density in \eqref{eq:mathcalP2}. We emphasize that $P_1$ is evaluated at $x=-L/2^-$ in the definition of $\mathcal{P}_1(-L/2^-)$. 
We further define
\begin{align*}
    \mathcal{P}(L/2)
    =p_0 \mathcal{P}_0(L/2)
    +p_1 \mathcal{P}_1(L/2),
\end{align*}
where 
\begin{align*}
    \mathcal{P}_i(L/2)
    &=\int_{a\Gamma} P_i(L/2,y,z)\,\mu(y,z)\,\dd y\,\dd z,\quad i\in\{0,1\}.
\end{align*}
Define
\begin{align*}
    \overline{P}_i(x)
    &:=\frac{1}{|a\Gamma|}\int_{a\Gamma}P_i(x,y,z)\,\dd y\,\dd z,\quad i\in\{0,1\},
\end{align*}
and $\overline{P}(x)=p_0\overline{P}_0(x)+p_1\overline{P}_1(x)$ and $\overline{Q}(x)=\overline{P}_0(x)-\overline{P}_1(x)$. 
Hence, integrating \eqref{eq:backward}, using the two-dimensional divergence theorem and the reflecting boundary conditions, we find that $\overline{P}_0$ and $\overline{P}_1$ satisfy the backward Kolmogorov equations \eqref{eq:backward} (which are now ordinary differential equations in the single variable $x$) with boundary condition
\begin{align}\label{eq:bc1}
    \overline{P}_1'(-L/2^+)
    =\overline{P}'(-L/2^+)-p_0\overline{Q}'(-L/2^+)
    &=0.
\end{align}
Therefore, we obtain $\overline{P}(x)$ and $\overline{Q}(x)$ (and therefore $\overline{P}_0(x)$ and $\overline{P}_1(x)$) as the following explicit functions of the three unknowns $\overline{P}(-L/2^+)$, $\overline{P}(L/2)$, and $\overline{Q}(L/2)$,
\begin{align}
\begin{split}\label{eq:PQexplicit}
    \overline{P}(x)
    &=\frac{x (\overline{P}(L/2)-\overline{P}(-L/2^+))}{L}+\frac{\overline{P}(-L/2^+)+\overline{P}(L/2)}{2},\quad x\in(-L/2,L/2],\\
    \overline{Q}(x)
    &=\text{sech}(\gamma_{\text{c}}) \Bigg[\frac{(\overline{P}(-L/2^+)-\overline{P}(L/2)) \sinh \left(\frac{\gamma_{\text{c}} (L-2 x)}{2 L}\right)}{\gamma_{\text{c}} p_0}\\
    &\quad+\overline{Q}(L/2) \cosh \bigg(\gamma_{\text{c}} \Big(\frac{x}{L}+\frac{1}{2}\Big)\bigg)\Bigg]\quad x\in(-L/2,L/2],
\end{split}    
\end{align}
where $\gamma_{\text{c}}=\sqrt{\lambda L^2/\Dc}$ compares the switching rate to the timescale of diffusion in the channel.

\subsection{Four exact equations}

We now derive four exact equations involving the eight probabilities $\mathcal{P}_0(\pm L/2)$, $\mathcal{P}_1(\pm L/2)$,  $\overline{P}_0(\pm L/2)$, and $\overline{P}_1(\pm L/2)$.
Integrating \eqref{eq:Pbackward} over the hemisphere $H_R$ in \eqref{eq:hemisphere} in the west bulk and using the divergence theorem yields
\begin{align*}
    0
    =\int_{H_R}\Delta P\,\dd \x
    =\int_{\partial H_R^+}\partial_{\mathbf{n}}P\,\dd S
    +\int_{a\Gamma}\partial_x P(-L/2^-,y,z)\,\dd y\,\dd z,
\end{align*}
where $\partial H_R^+$ denotes the curved part of the boundary of the hemisphere $H_R$ in \eqref{eq:boundaryofhemisphere}. 
The strong Markov property implies that $P$ has the following monopole decay at far-field\cite{lawley2026escape},
\begin{align*}
    P(\x)
    \sim \frac{C_0(a\Gamma)\mathcal{P}(-L/2^-)}{\|\x\|}\quad\text{as }\|\x\|\to\infty,\,x<-L/2,
\end{align*}
and therefore
\begin{align*}
    \int_{\partial H_R^+}\partial_{\mathbf{n}}P\,\dd S
    \to
    -2\pi aC_0(\Gamma)\mathcal{P}(-L/2^-)<0
    \quad\text{as }R\to\infty.
\end{align*}
Furthermore, the second continuity condition in \eqref{eq:contentrance} implies
\begin{align*}
    \int_{a\Gamma}\partial_x P(-L/2^-,y,z)\,\dd y\,\dd z
    &=p_0\int_{a\Gamma}\partial_x P_0(-L/2^-,y,z)\,\dd y\,\dd z\\
    &=p_0\Big(\frac{\Dc}{\Dl}\Big)^{\alphal}\int_{a\Gamma}\partial_x P_0(-L/2^+,y,z)\,\dd y\,\dd z\\
    &=p_0\Big(\frac{\Dc}{\Dl}\Big)^{\alphal}|\Gamma|a\frac{a}{L}L\overline{P}_0'(-L/2^+).
\end{align*}
Therefore, we have the following exact equation,
\begin{align}\label{eq:exact1}
    \rhol G(\Gamma)\mathcal{P}(-L/2^-)
    =p_0L\overline{P}_0'(-L/2^+)
    =L\overline{P}'(-L/2^+).
\end{align}
Similarly, integrating \eqref{eq:Qbackward} in the west bulk and using the divergence theorem yields
\begin{align*}
    0
    &=\lambda/\Dl\int_{x<-L/2} Q\,\dd \x
    -\int_{a\Gamma}\partial_x Q(L/2^-,y,z)\,\dd y\,\dd z\\
    &=\lambda/\Dl\int_{x<-L/2} Q\,\dd \x
    -\int_{a\Gamma}\partial_x P_0(L/2^-,y,z)\,\dd y\,\dd z\\
    &=\lambda/\Dl\int_{x<-L/2} Q\,\dd \x
    -\frac{2a\pi C_0(\Gamma)\mathcal{P}(-L/2^-)}{p_0},
\end{align*}
and therefore we obtain the exact equation,
\begin{align}\label{eq:Qexact}
    G(\Gamma)\mathcal{P}(-L/2^-)
    =p_0\frac{\lambda}{\Dl}\frac{1}{a}\frac{1}{|\Gamma|}\int_{x<-L/2} Q\,\dd \x
    =p_0({\gb})^2\frac{1}{a}\frac{1}{a^2|\Gamma|}\int_{x<-L/2} Q\,\dd \x,
\end{align}
where $\gb=\sqrt{\lambda a^2/\Dl}$ and  $\int_{x<-L/2} Q\,\dd \x$ denotes the volume integral of $Q(x,y,z)$ over the entire west bulk.

Integrating \eqref{eq:Pbackward} in the east bulk and using the divergence theorem, the far-field behavior, and the continuity conditions in \eqref{eq:contentranceL} yields 
\begin{align*}
    0
    &=-2\pi aC_0(\Gamma)(1-\mathcal{P}(L/2))
    +\int_{a\Gamma}\partial_x P(L/2^+,y,z)\,\dd y\,\dd z\\
    &=-2\pi aC_0(\Gamma)(1-\mathcal{P}(L/2))
    +\Big(\frac{\Dc}{\Dr}\Big)^{\alphar}\int_{a\Gamma}\partial_x P(L/2^-,y,z)\,\dd y\,\dd z,
\end{align*}
and therefore we have the following exact equation,
\begin{align}\label{eq:exact3}
    L\overline{P}'(L/2^-)
    =G(\Gamma)\rhor(1-\mathcal{P}(L/2)).
\end{align}
Similarly, integrating \eqref{eq:Qbackward} and using the divergence theorem and the continuity conditions in \eqref{eq:contentranceL} yields
\begin{align*}
    0
    &=\lambda/\Dr\int_{x>L/2} Q\,\dd \x
    +\int_{a\Gamma}\partial_x Q(L/2^+,y,z)\,\dd y\,\dd z\\
    &=\lambda/\Dr\int_{x>L/2} Q\,\dd \x
    +\Big(\frac{\Dc}{\Dr}\Big)^{\alphar}\int_{a\Gamma}\partial_x Q(L/2^-,y,z)\,\dd y\,\dd z,
\end{align*}
and therefore we have the exact equation,
\begin{align}\label{eq:exact4}
    L\overline{Q}'(L/2^-)
    =\frac{-\lambda\rhor}{\Dr}\frac{1}{a|\Gamma|}\int_{x>L/2} Q\,\dd \x
    =-\rhor(\gbr)^2\frac{1}{a}\frac{1}{a^2|\Gamma|}\int_{x>L/2} Q\,\dd \x,
\end{align}
where $\gbr=\sqrt{\lambda a^2/\Dr}$.

\subsection{Approximations}

We now posit some approximations to turn the four exact equations in \eqref{eq:exact1}, \eqref{eq:Qexact}, \eqref{eq:exact3}, and \eqref{eq:exact4} into four approximate equations which will yield explicit estimates of the three quantities $\overline{P}(-L/2^+)$, $\overline{P}(L/2)$, and $\overline{Q}(L/2)$ in \eqref{eq:PQexplicit} and the fourth quantity $\mathcal{P}(-L/2^-)$ in \eqref{eq:mathcalPfull}. 
The approximations at $x=-L/2$ are
\begin{align}
    \mathcal{P}_0(-L/2)
    &\approx\overline{P}_0(-L/2),\label{eq:keyusual0}\\
    \mathcal{P}_1(-L/2^-)
    &\approx\overline{P}_1(-L/2^-),\label{eq:keyusual1}\\
    ({\gb})^2\frac{1}{a}\frac{1}{a^2|\Gamma|}\int_{x<-L/2} Q\,\dd \x
    &\approx(G(\Gamma)+{\gb})\overline{Q}(-L/2^-),\label{eq:Qintapprox}
\end{align}
and the analogous approximations at $x=L/2$ are
\begin{align}
    \mathcal{P}_0(L/2)
    &\approx\overline{P}_0(L/2),\label{eq:keyusual2}\\
    \mathcal{P}_1(L/2)
    &\approx\overline{P}_1(L/2),\label{eq:keyusual3}\\
    (\gbr)^2\frac{1}{a}\frac{1}{a^2|\Gamma|}\int_{x>L/2} Q\,\dd \x
    &\approx(G(\Gamma)+\gbr)\overline{Q}(L/2).\label{eq:Qintapprox2}
\end{align}
The approximations in \eqref{eq:keyusual0}-\eqref{eq:keyusual1} and \eqref{eq:keyusual2}-\eqref{eq:keyusual3} ignore variation in the channel cross-section (analogous to \eqref{eq:ignorecrosssectionalvariation}). The approximations in \eqref{eq:Qintapprox} and \eqref{eq:Qintapprox2} ignore variation in the channel cross-section and further use a simple approximation to the modified Helmholtz equation in \eqref{eq:Qbackward} which is valid for both small and large $\lambda$ (derived in Appendix B of Ref.~\citenum{lawley2026diffusive}).

Combining the approximation \eqref{eq:Qintapprox} with the exact equation \eqref{eq:Qexact} yields
\begin{align}\label{eq:anotherapprox}
    G(\Gamma)\mathcal{P}(-L/2^-)
    \approx p_0(G(\Gamma)+{\gb})\overline{Q}(-L/2^-).
\end{align}
Rearranging \eqref{eq:anotherapprox} and using \eqref{eq:keyusual0}-\eqref{eq:keyusual1} yields 
\begin{align}\label{eq:key}
    \mathcal{P}_1(-L/2^-)
    &\approx\frac{p_0\gamma_\text{b}}{G(\Gamma)+p_0\gamma_\text{b}}\mathcal{P}_0(-L/2),
\end{align}
which is equivalent to
\begin{align}\label{eq:equivniceform}
    \mathcal{P}(-L/2^-)
    &\approx \frac{1}{1+\frac{p_1}{p_0}\frac{G(\Gamma)}{G(\Gamma)+{\gb}}}\mathcal{P}_0(-L/2)
    =\Big(p_0+\frac{p_1p_0{\gb}}{G(\Gamma)+p_0{\gb}}\Big)\mathcal{P}_0(-L/2).
\end{align}
Hence, if we use the approximations \eqref{eq:keyusual0}-\eqref{eq:keyusual1} and \eqref{eq:key}, then the exact equation in \eqref{eq:exact1} becomes the following approximate boundary condition,
\begin{align}\label{eq:bc2}
    L\overline{P}_0'(-L/2^+)
    \approx\rhol G(\Gamma)\Big(1+\frac{p_1{\gb}}{G(\Gamma)+p_0{\gb}}\Big)\overline{P}_0(-L/2).
\end{align}

Furthermore, combining the approximations \eqref{eq:keyusual2}-\eqref{eq:keyusual3} with the exact equation \eqref{eq:exact3} yields the following approximate boundary condition,
\begin{align}\label{eq:bc3}
    L\overline{P}'(L/2^-)
    \approx G(\Gamma)\rhor(1-\overline{P}(L/2)).
\end{align}
Finally, combining the approximation in \eqref{eq:Qintapprox2} with the exact equation~\eqref{eq:exact4} yields the following approximate boundary condition,
\begin{align}\label{eq:bc4}
    L\overline{Q}'(L/2^-)
    &\approx-\rhor(G(\Gamma)+{\gbr})\overline{Q}(L/2).
\end{align}

\subsection{Solving the approximate one-dimensional problem}

If we treat the three approximate boundary conditions in \eqref{eq:bc2}, \eqref{eq:bc3}, and \eqref{eq:bc4} as equalities, then we obtain the following boundary value problem for $\widehat{P}(x)$ and $\widehat{Q}(x)$ which approximate $\overline{P}(x)$ and $\overline{Q}(x)$,
\begin{align*}
    \widehat{P}''
    &=0,\quad x\in(-L/2,L/2),\\
    \widehat{Q}''
    &=(\lambda/\Dc)\widehat{Q},\quad x\in(-L/2,L/2),\\
    \widehat{P}'
    &=p_0\widehat{Q}',\quad x=-L/2^+,\\
    L(\widehat{P}+p_1\widehat{Q})'
    &=\rhol G(\Gamma)\Big(1+\frac{p_1{\gb}}{G(\Gamma)+p_0{\gb}}\Big)(\widehat{P}+p_1\widehat{Q}),\quad x=-L/2^+,\\
    L\widehat{P}'
    &= \rhor G(\Gamma)(1-\widehat{P}),\quad x=L/2^-,\\
    L\widehat{Q}'
    &=-\rhor(G(\Gamma)+{\gbr})\widehat{Q}\quad x=L/2^-,
\end{align*}
where the first boundary condition is the exact boundary condition in \eqref{eq:bc1}. It is straightforward to solve this boundary value problem and obtain explicit formulas for $\widehat{P}(x)$ and $\widehat{Q}(x)$. For brevity, we omit these complicated formulas.

For our purposes, we are primarily interested in the following explicit approximation to $\mathcal{P}(-L/2^-)$ obtained from $\widehat{P}_0(-L/2)=\widehat{P}(-L/2^+)+p_1\widehat{Q}(-L/2^+)$ and \eqref{eq:equivniceform},
\begin{align}
    \mathcal{P}(-L/2^-)
    \approx\mathcal{P}_*
    &:=\Big(p_0+\frac{p_1p_0{\gb}}{G(\Gamma)+p_0{\gb}}\Big)\widehat{P}_0(-L/2)\nonumber\\
    &=\bigg[\frac{G(\Gamma)+p_0{\gb}}{G(\Gamma)p_0+p_0{\gb}}+G(\Gamma)\rhol (\mathcal{F}_{\text{c}}+\mathcal{F}_{\text{ce}})\bigg]^{-1},\label{eq:Pmain}
\end{align}
where $\mathcal{F}_{\text{c}}$ and $\mathcal{F}_{\text{ce}}$ are the following dimensionless factors,
\begin{align}
    \mathcal{F}_{\text{c}}
    &=1+\frac{p_1}{p_0}\frac{\tanh (\gc)}{\gc}>0,\label{eq:Fc}\\
    \mathcal{F}_{\text{ce}}
    &=\frac{1}{G(\Gamma) \rhor}+\frac{(p_1/p_0) \text{sech}^2(\gc)}{ \rhor (\gbr+G(\Gamma))+\gc  \tanh (\gc)}>0.\label{eq:Fce}
\end{align}
We note that 
\begin{align*}
    \mathcal{P}_*
    =\frac{\rhor}{\rhol}(1-\widehat{P}(L/2)),
\end{align*}
which means that, modulo the differences in diffusivity in the two bulks, a particle is equally likely to (i) go from far away in the west bulk to far away in the east bulk, as it is to (ii) go from far away in the east bulk to far away in the west bulk.

The formula $\mathcal{P}_*$ in \eqref{eq:Pmain} is our main result. In the next section, we explore its behavior in various parameter regimes, including showing that it is exact in certain regimes.

\section{\label{sec:asymptotics}Flux asymptotics}

By the relation in \eqref{eq:JP}, the analysis in section~\ref{sec:gating2} yields the following explicit approximation to the steady-state average flux of particles which enter the channel from far away in the west bulk and ultimately wander far into the east bulk,
\begin{align*}
    J_*
    :=(2\pi C_0(\Gamma)a\Dl c_\infty)\mathcal{P}_*,
\end{align*}
where $\mathcal{P}_*$ is the approximate splitting probability in \eqref{eq:Pmain}. We now explore the asymptotic predictions of this approximate flux formula by analyzing the behavior of $\mathcal{P}_*$ in various parameter regimes.

Recall that $\mathcal{P}_*$ in \eqref{eq:Pmain} is a function of six dimensionless parameters. Specifically, $\mathcal{P}_*$ depends on the fraction of time that the gate is open (denoted $p_0$), the three dimensionless switching rates which compare the switching rate to the timescales of diffusion in the west bulk, channel, and east bulk,
\begin{align*}
    \gb=\sqrt{\lambda a^2/\Dl},\quad
    \gc=\sqrt{\lambda L^2/\Dc},\quad
    \gbr=\sqrt{\lambda a^2/\Dr},
\end{align*}
and the following comparisons of the channel aspect ratio to the change in diffusivity at the west and east channel entrances,
\begin{align*}
    \rhol=\frac{L}{a}\Big(\frac{\Dl}{\Dc}\Big)^{\alphal},\quad
    \rhor=\frac{L}{a}\Big(\frac{\Dr}{\Dc}\Big)^{\alphar}.
\end{align*}

\subsection{Slow switching}

If the switching rate is much slower than the timescales of diffusion, then
\begin{align*}
    \lim_{\max\{\gb,\gbr,\gc\}\to0}\mathcal{P}_*
    =p_0\mathcal{P}_*^{\text{open}},
\end{align*}
where $\mathcal{P}_*^{\text{open}}$ is the ``always open'' splitting probability in \eqref{eq:Popen}. Hence, the naive estimate that the gated flux is simply $p_0$ multiplied by the ``always open'' flux holds for slow switching.

\subsection{Fast switching}

If the switching rate is much faster than the timescales of diffusion, then
\begin{align*}
    \lim_{\min\{\gb,\gbr,\gc\}\to\infty}\mathcal{P}_*
    =\mathcal{P}_*^{\text{open}},
\end{align*}
where $\mathcal{P}_*^{\text{open}}$ is in \eqref{eq:Popen}. Hence, fast switching is the same as always being open.

\subsection{Large $\rhor$}

In the large $\rhor=(L/a)(\Dr/\Dc)^{\alphar}$ limit, we have that $\lim_{\rhor\to\infty}\mathcal{F}_{\text{ce}}=0$, and therefore,
\begin{align*}
    \lim_{\rhor\to\infty}\mathcal{P}_*
    =\bigg[\frac{G(\Gamma)+p_0{\gb}}{G(\Gamma)p_0+p_0{\gb}}+G(\Gamma)\rhol \mathcal{F}_{\text{c}}\bigg]^{-1},
\end{align*}
which recovers the splitting probability derived in Ref.~\cite{lawley2026diffusive} for the case that particles are absorbed at the right end of the channel. This result is to be expected, since if $\rhor\gg1$, then any particle which reaches the east end of the channel will likely not return to the west end of the channel because either the channel is long (i.e.\ $L/a\gg1$) and/or the diffusivity is much faster in the east bulk than the channel and any non-It\^o multiplicative noise will force the particle out the right end of the channel (i.e.\ $(\Dr/\Dc)^{\alphar}\gg1$). 

\subsection{Large $\min\{\rhol,\rhor\}$}

If both $\rhol$ and $\rhor$ are large, then we claim that the splitting probability has the following exact asymptotic,
\begin{align}\begin{split}\label{eq:exactasymptotic}
    \mathcal{P}(-L/2^-)
    \sim\mathcal{P}_*
    &\sim\frac{1}{G(\Gamma)\rhol\mathcal{F}_{\text{c}}}\\
    &=\bigg[G(\Gamma)\Big(1+\frac{p_1}{p_0}\frac{\tanh (\gc)}{\gc}\Big)\frac{L}{a}\Big(\frac{\Dl}{\Dc}\Big)^{\alphal}\bigg]^{-1}\;\text{as }\min\{\rhol,\rhor\}\to\infty,
\end{split}    
\end{align}
where $\mathcal{P}(-L/2^-)$ is the exact splitting probability in \eqref{eq:mathcalPfull} (which is, in general, not known explicitly). In other words, the approximation $\mathcal{P}_*$ in \eqref{eq:Pmain} is exact if $\min\{\rhol,\rhor\}\gg1$. Since $\rhol=(L/a)(\Dl/\Dc)^{\alphal}$ and $\rhor=(L/a)(\Dr/\Dc)^{\alphar}$, we may have that $\min\{\rhol,\rhor\}\gg1$ if the channel is long (i.e.\ $L/a\gg1$) and/or the diffusivity is slow in the channel and we have non-It\^o interpretations of the multiplicative noise (i.e.\ $(\Dl/\Dc)^{\alphal}\gg1$ and $(\Dr/\Dc)^{\alphar}\gg1$).

The second asymptotic equivalence and the final equality in \eqref{eq:exactasymptotic} follow immediately from \eqref{eq:Pmain}-\eqref{eq:Fce}. The first asymptotic equivalence follows from the strong Markov property and results in Refs.~\cite{richardson2026yet, lawley2026diffusive}. Before detailing the argument, we sketch the main idea. Specifically, it was shown in Ref.~\cite{richardson2026yet} that if $\rhor\gg1$, then a particle which starts at the east channel entrance is unlikely to reach the west channel entrance. Therefore, the probability of wandering far into the east bulk starting from the west channel entrance is the same as the probability of merely reaching the east channel entrance from the west channel entrance, and the asymptotic of this probability for $\rhol\gg1$ was found in Ref.~\cite{lawley2026diffusive}. 

More precisely, the strong Markov property implies that
\begin{align}\label{eq:squeeze}
\begin{split}
&\Big(\int_{a\Gamma}h(-L/2^-,y,z)\,\mu(y,z)\,\dd y\,\dd z\Big)\inf_{(y,z)\in a\Gamma,\,i\in\{0,1\}}P_i(L/2,y,z)
\le\mathcal{P}(-L/2^-)\\
    &\quad\le \Big(\int_{a\Gamma}h(-L/2^-,y,z)\,\mu(y,z)\,\dd y\,\dd z\Big)\sup_{(y,z)\in a\Gamma,\,i\in\{0,1\}}P_i(L/2,y,z),
\end{split}    
\end{align}
where $h(x,y,z)$ is the probability that a particle starting at $(x,y,z)$ will reach the east end of the channel given that the gate is initially in its stationary distribution. In words, \eqref{eq:squeeze} says that in order to go far into the east bulk from the west channel entrance, the particle must first reach the east channel entrance (which happens with probability given by the integral of $h$ in \eqref{eq:squeeze}), and then the particle must go far into the east bulk from the east channel entrance (which is bounded below and above by the infimum and supremum terms in \eqref{eq:squeeze}).

It was shown in Ref.~\cite{lawley2026diffusive} that the integral of $h$ in \eqref{eq:squeeze} has the following asymptotic behavior,
\begin{align}\label{eq:firstpart}
    \int_{a\Gamma}h(-L/2^-,y,z)\,\mu(y,z)\,\dd y\,\dd z
    \sim\frac{1}{G(\Gamma)\rhol\mathcal{F}_{\text{c}}}\quad\text{as }\rhol\to\infty.
\end{align}
Furthermore,
\begin{align*}
    1-P_i(L/2,y,z)\le u(L/2,y,z)\quad\text{for all }(y,z)\in a\Gamma,\,i\in\{0,1\},
\end{align*}
where $u(x,y,z)$ is the probability that a particle starting at $(x,y,z)$ will reach the west end of the channel. It was shown in Ref.~\cite{richardson2026yet} that
\begin{align*}
    \lim_{\rhor\to\infty}u(L/2,y,z)=0\quad\text{for all }(y,z)\in a\Gamma,
\end{align*}
and therefore
\begin{align}\label{eq:secondpart}
    \lim_{\rhor\to\infty}P_i(L/2,y,z)=1\quad\text{for all }(y,z)\in a\Gamma,\,i\in\{0,1\}.
\end{align}
Hence, combining \eqref{eq:firstpart} and \eqref{eq:secondpart} with \eqref{eq:squeeze} and using the squeeze theorem yields
\begin{align*}
    \mathcal{P}(-L/2^-)
    \sim\frac{1}{G(\Gamma)\rhol\mathcal{F}_{\text{c}}}\quad\text{as }\min\{\rhol,\rhor\}\to\infty,
\end{align*}
which yields \eqref{eq:exactasymptotic}.

\subsection{\label{sec:smallrhol}Small $\rhol$}

In the limit that $\rhol=(L/a)(\Dl/\Dc)^{\alphal}$ is small, we obtain
\begin{align}\label{eq:Pgamma}
    \lim_{\rhol\to0}\mathcal{P}_*
    =p_0+\frac{p_1p_0\gb}{G(\Gamma)+p_0\gb}
    =p_0\Big(\frac{G(\Gamma)+\gb}{G(\Gamma)+p_0\gb}\Big)
    =:\mathcal{P}_{*}^{\text{gated $\Gamma$}},
\end{align}
where the final equality defines $\mathcal{P}_{*}^{\text{gated $\Gamma$}}$. We note that \eqref{eq:Pgamma} is not exact in the sense that for a general $\Gamma\subset\R^2$,
\begin{align}\label{eq:Pnotexact}
    \lim_{\rhol\to0}\mathcal{P}(-L/2^-)\neq \mathcal{P}_{*}^{\text{gated $\Gamma$}},
\end{align}
where $\mathcal{P}(-L/2^-)$ is the exact splitting probability in \eqref{eq:mathcalPfull} (which is not known explicitly).

However, \eqref{eq:Pnotexact} is to be expected. To see this, observe first that $\lim_{\rhol\to0}\mathcal{P}(-L/2^-)$ is the probability that a diffusing particle will hit the west entrance to the channel when the gate is open, since if $\rhol\to0$ and all other parameters are fixed, then any particle which reaches the west entrance to the channel when it is open will enter the channel and not exit at the west end of the channel since $\rhol\ll\rhor$, and will therefore eventually wander far away into the east bulk. Hence, $\lim_{\rhol\to0}\mathcal{P}(-L/2^-)$ is the probability that a particle will get absorbed at a stochastically gated target with shape $\Gamma$, which is notoriously difficult to calculate, even if $\Gamma$ is a disk \cite{szabo1982}.

Though $\mathcal{P}_{*}^{\text{gated $\Gamma$}}$ is not exact in the sense of \eqref{eq:Pnotexact}, the stochastic simulations in section~\ref{sec:shortchannelsims} below show that $\mathcal{P}_{*}^{\text{gated $\Gamma$}}$ is a very accurate approximation of $\lim_{\rhol\to0}\mathcal{P}(-L/2^-)$, at least in the case that $\Gamma$ is a disk (Figure~\ref{fig:shortchannel}A) or $\Gamma$ is a square (Figure~\ref{fig:shortchannel}B).

\subsection{Short channel}

To understand the behavior of $\mathcal{P}_*$ in the case that the channel is short (i.e.\ $L/a\ll1$), we introduce a small dimensionless parameter $0<\eps\ll1$ and set
\begin{align*}
    \rhol=\eps(\Dl/\Dc)^{\alphal},\quad
    \rhor=\eps(\Dr/\Dc)^{\alphar},
\end{align*}
and suppose $\gc=\sqrt{\lambda L^2/\Dc}=\mathcal{O}(\eps)$ as $\eps\to0$. In this case, 
\begin{align}\label{eq:intuitive}
    \lim_{\eps\to0}\mathcal{P}_*
    =\frac{(\Dr/\Dc)^{\alphar}}{(\Dl/\Dc)^{\alphal}+(\Dr/\Dc)^{\alphar}}\mathcal{P}_{*}^{\text{gated $\Gamma$}},
\end{align}
where $\mathcal{P}_{*}^{\text{gated $\Gamma$}}$ is defined in \eqref{eq:Pgamma}.

The result \eqref{eq:intuitive} is intuitive because in order to wander far away into the east bulk, the particle must first reach the west entrance to the channel when it is open (which occurs with approximate probability $\mathcal{P}_{*}^{\text{gated $\Gamma$}}$, see section~\ref{sec:smallrhol} above) and then the particle must wander far away into the east bulk rather than the west bulk. If $\Dl=\Dr$ and $\alphal=\alphar$ or $\alphal=\alphar=0$, then the prefactor in \eqref{eq:intuitive} is 1/2, which accords with symmetry. If $(\Dr/\Dc)^{\alphar}>(\Dl/\Dc)^{\alphal}$, then the prefactor is larger than 1/2, which reflects the bias for the particle to wander into the east bulk (owing to the spurious force induced by any non-It\^o convention \cite{serov2020statistical}).

\subsection{Fast and slow switching in bulks versus channel} 

The fact that there are three diffusivities ($\Dl$, $\Dc$, and $\Dr$) and two lengthscales ($L$ and $a$) means that the effective switching rates can be very different in the west bulk versus the channel versus the east bulk. For example, it is possible to have
\begin{align*}
    \gamma_{i}
    \ll\gamma_{j}
    \ll\gamma_{k},
\end{align*}
for any permutation of the subscripts $i,j,k$ in $\{\text{w},\text{c},\text{e}\}$.

\begin{figure}[t]
\centering
\includegraphics[width=.6\linewidth]{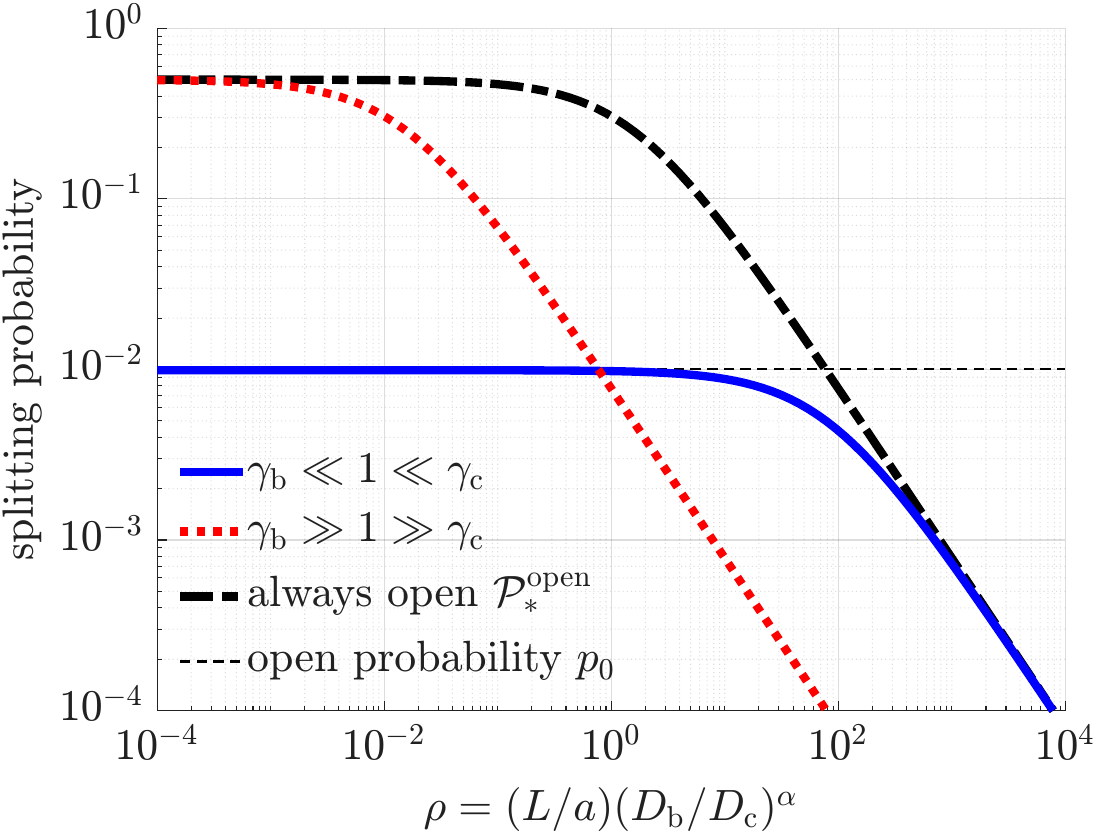}
\caption{Comparison of the limiting splitting probabilities in \eqref{eq:fastchannelswitching}-\eqref{eq:fastbulkswitching}.}
\label{fig:fastslow}
\end{figure}

To investigate some of these cases, suppose that the two bulks have identical diffusivity $\Dl=\Dr=\Db$ and $\alphal=\alphar=\alpha$ so that $\gb=\gbr=\gamma_{\text{b}}:=\sqrt{\lambda a^2/\Db}$ and $\rhol=\rhor=\rho:=(L/a)(\Db/\Dc)^\alpha$. In the limit that the effective bulk switching rate is slow and the effective channel switching rate is fast, we have that 
\begin{align}\label{eq:fastchannelswitching}
    \lim_{\gamma_{\text{b}}\to0,\gc\to\infty}\mathcal{P}_*
    =\frac{p_0 }{1+p_0+p_0G(\Gamma) \rho}.
\end{align}
In the limit that the effective bulk switching rate is fast and the effective channel switching rate is slow, we have that 
\begin{align}\label{eq:fastbulkswitching}
    \lim_{\gamma_{\text{b}}\to\infty,\gc\to0}\mathcal{P}_*
    =\frac{p_0 }{2p_0+G(\Gamma)\rho}.
\end{align}
Interestingly, \eqref{eq:fastchannelswitching}-\eqref{eq:fastbulkswitching} imply the following ordering for small $\rho$,
\begin{align*}
    \lim_{\gamma_{\text{b}}\to0,\gc\to\infty}\mathcal{P}_*
    <\lim_{\gamma_{\text{b}}\to\infty,\gc\to0}\mathcal{P}_*\quad\text{if }\rho<1/G(\Gamma),
\end{align*}
whereas the reverse holds for large $\rho$,
\begin{align*}
    \lim_{\gamma_{\text{b}}\to0,\gc\to\infty}\mathcal{P}_*
    >\lim_{\gamma_{\text{b}}\to\infty,\gc\to0}\mathcal{P}_*\quad\text{if }\rho>1/G(\Gamma).
\end{align*} 
This result is illustrated in Figure~\ref{fig:fastslow}, where we plot the limiting expressions in \eqref{eq:fastchannelswitching}-\eqref{eq:fastbulkswitching} as a function of $\rho$ for the case that $p_0=1\%$ and $\Gamma$ is a disk.

\section{\label{sec:simulations}Stochastic simulations}

We now compare the approximate splitting probability $\mathcal{P}_*$ in \eqref{eq:Pmain} to stochastic simulations in the case that the two bulks have identical diffusivity $\Dl=\Dr=\Db$ and $\alphal=\alphar=\alpha$ so that $\gb=\gbr=\gamma_{\text{b}}=\sqrt{\lambda a^2/\Db}$. The simulation algorithms are described in Appendix~\ref{sec:algorithm}. 

\subsection{\label{sec:shortchannelsims}Short channel limit}

\begin{figure}[t]
\centering
\includegraphics[width=1\linewidth]{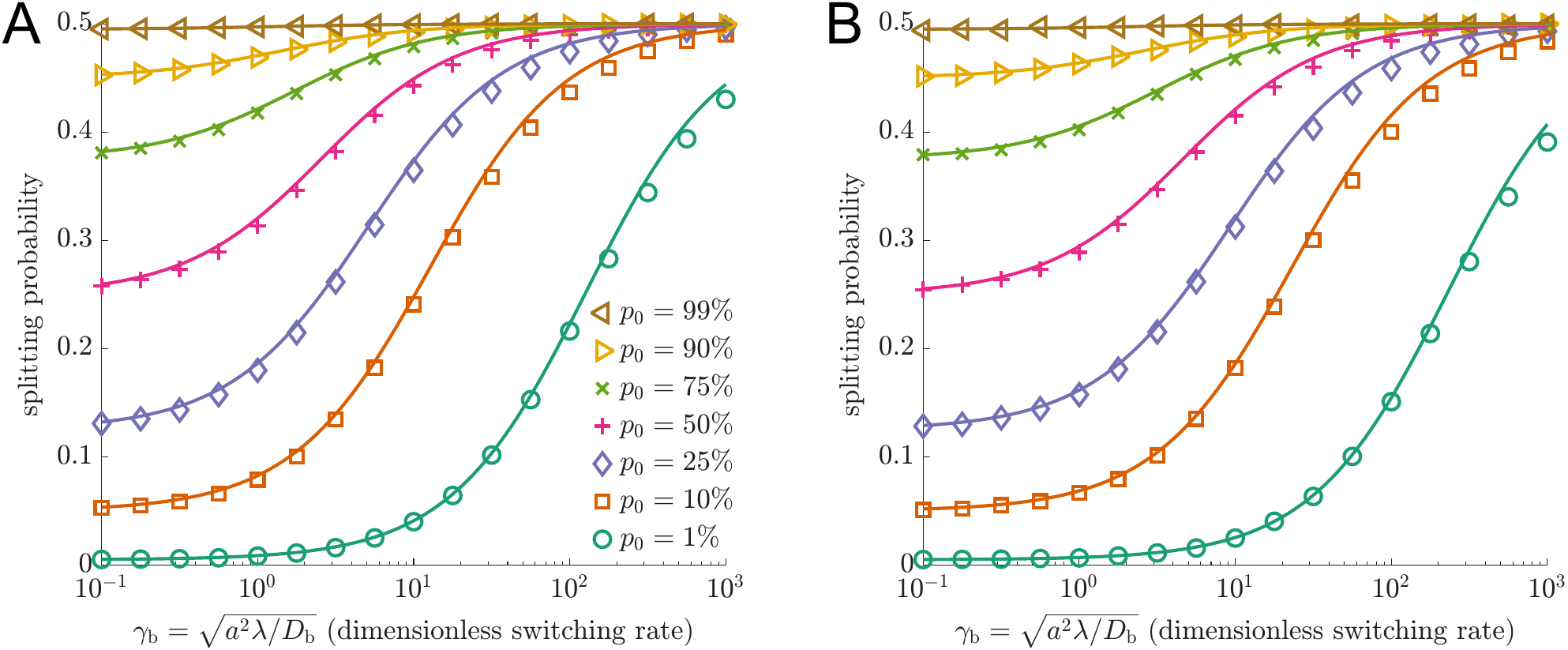}
\caption{Comparison of the short channel formula in \eqref{eq:shortchannelformula} (curves) to stochastic simulation (markers) for different values of the open probability $p_0$. In Panel A, $\Gamma$ is the unit disk. In Panel B, $\Gamma$ is the unit square.}
\label{fig:shortchannel}
\end{figure}

In the short channel limit, we have that 
\begin{align}\label{eq:shortchannelformula}
    \lim_{L\to0}\mathcal{P}_*
    =\frac{p_0}{2}\Big(\frac{G(\Gamma)+\gamma_{\text{b}}}{G(\Gamma)+p_0\gamma_{\text{b}}}\Big)
    =\frac{\mathcal{P}_*^{\text{gated $\Gamma$}}}{2},
\end{align}
where $\mathcal{P}_*^{\text{gated $\Gamma$}}$ is \eqref{eq:Pgamma}. 
In Figure~\ref{fig:shortchannel}, we plot the formula in \eqref{eq:shortchannelformula} as a function of $\gamma_{\text{b}}$ for different values of the open probability $p_0$ (curves), which compares favorably to stochastic simulations (markers). Indeed, Figure~\ref{fig:shortchannel} shows that $\mathcal{P}_*$ is an accurate approximation (though not exact) in the regime of \eqref{eq:shortchannelformula}. 
Figure~\ref{fig:shortchannel}A is for the case that $\Gamma$ is the unit disk, and thus $G(\Gamma)=4/\pi$. Figure~\ref{fig:shortchannel}B is for the case that $\Gamma$ is the unit square, and thus $G(\Gamma)=2\pi C_0(\Gamma)/|\Gamma|\approx2.31$ since the electrostatic capacitance of the unit square is $C_0(\Gamma)\approx0.367$ \cite{read1997improved}.

\subsection{Variable channel length}

\begin{figure*}[t]
\centering
\includegraphics[width=1\linewidth]{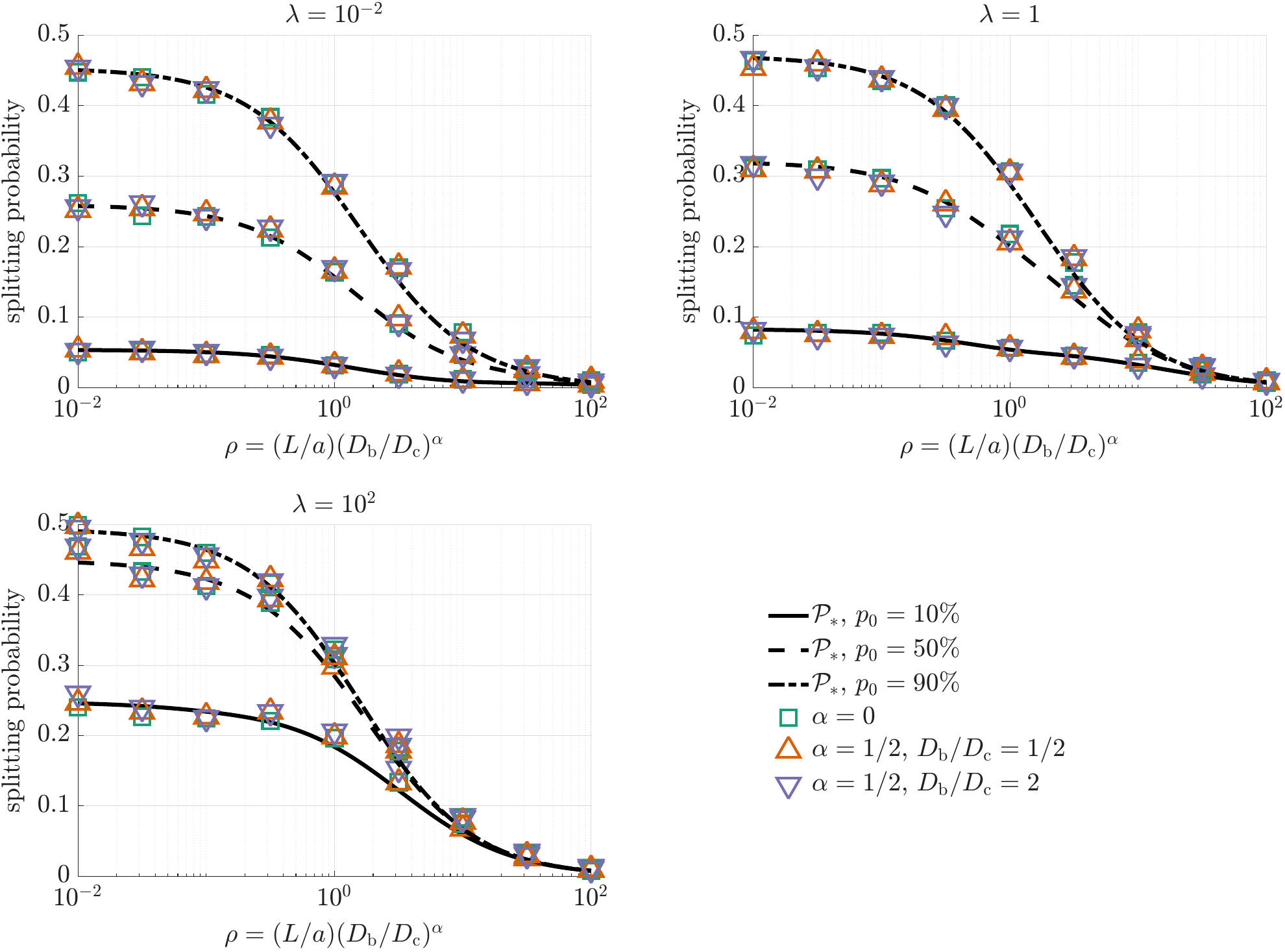}
\caption{Comparison of $\mathcal{P}_{*}$ in \eqref{eq:Pmain} (black curves) to stochastic simulations (markers).}
\label{fig:traverse}
\end{figure*}

Figure~\ref{fig:traverse} plots $\mathcal{P}_*$ as a function of $\rho=(L/a)(\Db/\Dc)^\alpha$ for different values of the switching rate $\lambda$ (different panels) and the open probability $p_0$. This figure shows that the approximation agrees well with the stochastic simulations (markers). The channel is cylindrical in this plot (i.e.\ $\Gamma$ is the unit disk and $G(\Gamma)=4/\pi$).

\section{\label{sec:discussion}Discussion}

In this paper, we used partial differential equation analysis to estimate the average diffusive flux through a channel. Our explicit flux estimate accounts for stochastic gating, channel geometry, and differing diffusivity inside versus outside the channel. We showed that our flux estimate is exact in some parameter regimes, and we used stochastic simulations to show that it remains accurate across other parameter regimes.

Perhaps the first mathematical analysis of the diffusive flux through a stochastically gated channel is the student thesis \cite{lawleythesis} published later in \cite{lawley15sima}. Refs.~\cite{lawleythesis, lawley15sima} were motivated by insect respiration, in which the flow of oxygen and carbon dioxide through tracheal tubes is modulated by a rapidly opening and closing gate called a spiracle \cite{wigglesworth31}. The mathematical methods developed in Refs.~\cite{lawleythesis, lawley15sima} were used to explain certain aspects of the discontinuous gas exchange cycle in insects \cite{lawley2020spiracular, lawley2022water}. Alternative mathematical approaches to the flux through a stochastic gate were developed in several other works \cite{bressloff2015moment, lawley2016boundary}, including Ref.~\cite{PB5} which accounted for branching channels (motivated by the branched structure of insect tracheae), Ref.~\cite{lawley2019electrodiffusive} which used the Nernst--Planck equation to model the movement of ions in response to a concentration gradient and an electric
field, and Ref.~\cite{bressloff_diffusion_2018} which considered non-Markovian gating.

We conclude by comparing our flux estimate to an estimate derived in Ref.~\cite{berezhkovskii2017effect} and studied in Refs.~\cite{berezhkovskii2018effect, berezhkovskii2018stochastic}. Specifically, the following flux estimate was posited in Ref.~\cite{berezhkovskii2017effect} for the case of a cylindrical channel between two bulks with identical diffusivities $\Dl=\Dr=\Db$, (see Equation~(3.4) in Ref.~\citenum{berezhkovskii2017effect}),
\begin{align}\label{eq:J17}
    J_{2017}
    &=\frac{4a\Db c_\infty}{2+\nu+\frac{p_1}{p_0}\frac{2\gc\nu+(\gc^2+\nu^2)\tanh(\gc)}{\gc(\nu+\gc\tanh(\gc))}},\quad \text{where }\nu=\frac{4}{\pi}\frac{L}{a}\frac{\Db}{\Dc}.
\end{align}
One discrepancy between $J_{2017}$ in \eqref{eq:J17} and our estimate $J_*$ in \eqref{eq:J0} is that $J_{2017}$ does not depend on the multiplicative noise convention (i.e.\ there is no $\alpha$ in $J_{2017}$). This discrepancy stems from the derivation of $J_{2017}$ which employed a certain effective escape rate from a channel \cite{bezrukov2000particle} which was recently shown to be valid for $\alpha=1$ or $\Db=\Dc$ \cite{lawley2026escape}. 

However, significant discrepancies between $J_{2017}$ and $J_*$ persist even if we set $\alpha=1$ in $J_*$ or take $\Db=\Dc$. For example, $J_{2017}$ does not approach the ``always open'' flux for fast gating,
\begin{align*}
    \lim_{\lambda\to\infty}J_{2017}
    \neq \lim_{p_0\to1}J_{2017}, 
\end{align*}
which contrasts $J_*$,
\begin{align*}
    \lim_{\lambda\to\infty}J_*
    =\lim_{p_0\to1}J_*.
\end{align*}
Furthermore, $J_*$ is independent of $\lambda$ if the channel is short,
\begin{align}\label{eq:short2017}
    \lim_{L\to0}J_{2017}
    =4a\Db c_\infty \Big(\frac{p_0}{2}\Big),
\end{align}
which contrasts $J_*$,
\begin{align}\label{eq:shortstar}
    \lim_{L\to0}J_*
    =4a\Db c_\infty \Big(\frac{p_0}{2}\Big)\Big(\frac{4/\pi+\gamma_{\text{b}}}{4/\pi+p_0\gamma_{\text{b}}}\Big),\quad \text{where }\gamma_{\text{b}}=\sqrt{\lambda a^2/\Db}.
\end{align}
The predictions in \eqref{eq:short2017}-\eqref{eq:shortstar} differ sharply if switching is fast compared to the timescale of diffusion in the bulk. Indeed,
\begin{align*}
    \lim_{\gamma_{\text{b}}\to\infty}\Big(\frac{\lim_{L\to0}J_{2017}}{\lim_{L\to0}J_{*}}\Big)
    =p_0,
\end{align*}
which predicts that $J_{2017}$ underestimates the actual flux. 
The accuracy of \eqref{eq:shortstar} is supported by Figure~\ref{fig:shortchannel}A. 


\section*{Acknowledgments}

The author gratefully acknowledges support from the National Science Foundation (NSF DMS-2325258).

\section*{Data availability}

The data that support the findings of this study are available from the corresponding author upon reasonable request.

\appendix

\section{\label{sec:algorithm}Stochastic simulation algorithms}

\subsection{Short channel simulation algorithm}

We start by describing the stochastic simulation algorithm used to produce the markers in Figure~\ref{fig:shortchannel}. Each marker in Figure~\ref{fig:shortchannel}A is computed from $5\times10^6$ independent simulations, and each marker in Figure~\ref{fig:shortchannel}B is computed from $8\times10^5$ independent simulations

Owing to symmetry, the simulations in section~\ref{sec:shortchannelsims} require merely (1) calculating the probability that a particle which reaches the west channel entrance from far away in the west bulk will eventually hit this west channel entrance when the gate is open and (2) multiplying this probability by $1/2$. For the case that the channel cross-section is a disk (i.e.\ Figure~\ref{fig:shortchannel}A), the simulation of (1) follows a simulation algorithm used in Ref.~\cite{lawley2026diffusive}. For the case that the channel cross-section is a square (i.e.\ Figure~\ref{fig:shortchannel}B), we follow the following modification of that simulation algorithm.

\textbf{Step 0.} Initialize the gate to be open with probability $p_0$ and closed with probability $p_1=1-p_0$. If the gate is initially open, then the particle is absorbed and the simulation ends. If the gate is initially closed, then we must initialize the particle to be at a random position on the unit square,
\begin{align}\label{eq:initial}
    \X(0)
    =(0,Y(0),Z(0))\in\{0\}\times[-1/2,1/2]^2,
\end{align}
where the distribution of $(Y(0),Z(0))$ is the harmonic measure $\mu$ in \eqref{eq:mathcalPfull0}-\eqref{eq:mathcalPfull1}. While $\mu$ is known explicitly if $\Gamma$ is a disk (see \eqref{eq:hittingdensity}), we do not know an explicit formula for $\mu$ if $\Gamma$ is a square. Nevertheless, one can approximate samples of $\mu$ on the square $\Gamma$ by first letting $(X',Y',Z')$ be uniformly distributed on any sphere containing the unit disk, and then following the algorithm of Bernoff, Lindsay, and Schmidt \cite{bernoff2018boundary} until the particle either (i) reaches $\Gamma$ or (ii) reaches a large distance $R_\infty=10^{10}$ from the origin. If (i) happens, then the resulting position on the square gives \eqref{eq:initial}. If (ii) happens, then we repeat until (i) happens.

Proceed to Step 1.

\textbf{Step 1.} Given the particle $\X_0=(0,Y_0,Z_0)\in \Gamma$ and given that the gate is closed, we sample a realization of the waiting time until the gate opens, which is given exactly by the following exponential random variable with mean $1/\lambda_1$,
\begin{align*}
    \tau
    =E/\lambda_1,
\end{align*}
where $E$ is a unit mean exponential random variable. Before time $\tau$ elapses, the particle diffuses in three dimensions with a reflecting boundary condition at $x=0$, and hence its position when the gate opens is given exactly by
\begin{align}
    X
    &=-|0+\sqrt{2\Db\tau}\mathcal{N}_x|,\label{eq:X}\\
    Y
    &=Y_0+\sqrt{2\Db\tau}\mathcal{N}_y,\label{eq:Y}\\
    Z
    &=Z_0+\sqrt{2\Db\tau}\mathcal{N}_z,\label{eq:Z}
\end{align}
where $\mathcal{N}_x,\mathcal{N}_y,\mathcal{N}_z,$ are independent standard normal random variables.

\textbf{Step 2:} Starting from $\X=(X,Y,Z)$ in \eqref{eq:X}-\eqref{eq:Z}, we follow the algorithm in Ref.~\citenum{bernoff2018boundary} to simulate the particle trajectory until the particle either (i) reaches a distance $R_\infty=10^{10}\gg1$ from the west channel entrance or (ii) hits the west channel entrance. If (i) occurs, then we assume that the particle will never return to the channel and the simulation ends. If (ii) occurs, then we set the gate to be closed with probability 
\begin{align}\label{eq:closedsimulation}
    p_1(1-e^{-(\lambda_0+\lambda_1)t}),
\end{align}
where $t$ is the time elapsed since the start of Step 2. If the gate is open, then the particle is absorbed and the simulation ends. If the gate is closed, then return to Step 1.

\subsection{Variable channel length simulations}

We now describe the simulation algorithm used to produce the markers in Figure~\ref{fig:traverse}. Each marker in Figure~\ref{fig:traverse} is computed from $8\times10^3$ independent simulations.

\textbf{Step 1.} Follow a simulation algorithm used in Ref.~\cite{lawley2026diffusive} to simulate the particle until it either (i) wanders a large distance $R_\infty=10^{10}$ away from the channel in the west bulk or (ii) reaches the east channel entrance. If (i) happens, then it is assumed that the particle will never wander far away into the east bulk and the simulation ends. If (ii) happens, then proceed to Step 2. 

\textbf{Step 2.} Follow a simulation algorithm used in Ref.~\cite{lawley2026diffusive} to simulate the particle until it either (i) wanders a large distance $R_\infty=10^{10}$ away from the channel in the east bulk or (ii) reaches the west channel entrance (where the gate is). If (i) happens, then it is assumed that the particle has wandered far away into the east bulk and will never return to the west bulk and the simulation ends. If (ii) happens, then return to Step 1.

\bibliography{library}
\bibliographystyle{abbrv}       

\end{document}